# VIZSLA – Versatile Ice Zigzag Sublimation Setup
# for Laboratory Astrochemistry

Gábor Bazsó,[a] István Pál Csonka,[b] Sándor Góbi,[b] and György Tarczay[b,c*]

[a] Wigner Research Centre for Physics, PO Box 49, H-1525 Budapest, Hungary

[b] MTA-ELTE Lendület Laboratory Astrochemistry Research Group, Institute of Chemistry,
ELTE Eötvös Loránd University, H–1518 Budapest, Hungary

[c] Laboratory of Molecular Spectroscopy, Institute of Chemistry, ELTE Eötvös Loránd
University, H–1518 Budapest, Hungary

*corresponding author, e-mail: gyorgy.tarczay@ttk.elte.hu

## Abstract

In this article a new, multi-functional, high-vacuum astrophysical ice setup, VIZSLA (Versatile
Ice Zigzag Sublimation Setup for Laboratory Astrochemistry), is introduced. The instrument
allows the investigation of astrophysical processes both in a low-temperature $para$-$H_2$ matrix
and in astrophysical analog ices. In $para$-$H_2$ matrix the reaction of astrochemical molecules
with H atoms and $H^+$ ions can be studied very effectively. For the investigation of astrophysical
analog ices the setup is equipped with different irradiation and particle sources: an electron gun,
for modeling cosmic rays; an H atom beam source (HABS); a microwave H atom lamp, for
generating H Lyman-$\alpha$ radiation, and a tunable (213 nm to 2800 nm) laser source. For analysis,
an FT-IR (and a UV-Visible) spectrometer and a quadrupole mass analyzer are available. The
setup has two cryostats, offering novel features for analysis. Upon the so-called temperature-
programmed desorption (TPD) the molecules, desorbing from the first cryostat, can be mixed
with Ar and can be deposited onto the substrate of the other cryostat. The well-resolved
spectrum of the molecules isolated in an Ar matrix serves a unique opportunity to identify the
desorbing products of a processed ice. Some examples are provided to show how the $para$-$H_2$
matrix experiments and the TPD – matrix-isolation recondensation experiments can help to
understand astrophysically important chemical processes at a low temperature. It is also
discussed, how these experiments can complement the studies carried out by similar
astrophysical ice setups.

# I. INTRODUCTION

It was believed in the early 20[th] century that, due to radiations, chemical species such as molecules or radicals cannot exist in the interstellar medium (ISM). The first prediction for the presence of a diatomic species, CH, in the ISM was made in the late 1930s.[1] This radical was identified in the ISM only 3 years later based on the comparison of spectra recorded in the laboratory and measured by telescopes.[2] Since the start of Atacama Large Millimeter/submillimeter Array (ALMA) observations in 2011, the number of identified chemical species in the ISM has largely increased, nowadays, not counting isotopomers, about 250 molecular species have been unambiguously identified.[3,4] It can be expected that future missions, *e.g.*, the James Webb Space Telescope (JWST),[5] will also have a large impact on observational astrochemistry and will further increase the diversity of identified astrochemical species.

The solid identification of the chemical species in the ISM can only be based on the comparison of laboratory experiments and telescopic observations. Furthermore, the understanding of the formation and destruction of molecules in different celestial objects also requires laboratory investigations. Since most of the astrophysical environments are extreme compared to terrestrial conditions, it is challenging to model these conditions in the laboratory, and it requires special setups dedicated to laboratory astrochemistry experiments. The chemistry undergoing in these conditions can also be fundamentally different from that learned in a standard organic or inorganic laboratory.

The early laboratory astrochemistry focused mostly on gas-phase studies. Now it is well accepted that for understanding the composition of the gaseous ISM, it is inevitable to study solid-state astrochemical processes.[6] Both theoretical modeling and laboratory experiments proved that there are a number of important and abundant interstellar species, which can only form on the icy surface of interstellar grains, and are released into the gaseous ISM by desorption. Among these species, the most important is the hydrogen molecule ($H_2$).[7,8]

In a typical solid-state astrophysical ice experiment, small molecules are deposited onto a cryogenic surface in an ultrahigh vacuum (UHV, $1 \cdot 10^{-8}$ mbar – $10^{-11}$ mbar) chamber. The reactions are triggered either by different types of radiations, *e.g.,* UV-Vis, Hydrogen Lyman-α, or X-rays, or by electron beam (mimicking the effect of cosmic rays), by ion bombardment or by reactive atoms, including e.g. H, C, or N atoms.[9,10] The reaction is spectroscopically monitored in the solid-state and usually the species subliming upon a temperature-programmed desorption (TPD) process are also analyzed, most often by mass spectrometry. Far from a complete list, some of the most well-known special setups designed for these kinds of studies include the Keck Machine (at University of Hawaii, USA),[11] CRYOPAD and SURFRESIDE[3] (at University of Leiden, Netherland),[9,10] FORMOLISM and VENUS (at CY Cergy Paris University, France),[12,13] ICA (at Institute of Nuclear Research, Debrecen, Hungary),[14] SPACE CAT[2] (at Harvard University),[15] PAC (at Open University),[16] the STARK chamber (at the Universidade do Vale do Paraíba),[17] as well as further cryogenic setups at I[3]OLAB (NASA AMES, USA),[18] at the Osservatorio Astrofisico di Catania,[19] at the Laboratory Astrophysics Group at the Friedrich Schiller University Jena,[20] and at the Astrophysical Chemistry / Ice & Planetary Science Group at Hokkaido University, Japan.[21]

Generally, the deposited and the processed ices are investigated in the solid phase by IR, Raman, and UV-VIS spectroscopic methods, while a quadrupole mass spectrometer (QMS) with electron impact ionization is used for the analysis of the subliming molecules. The main difficulty of the solid-phase spectroscopic investigation of the ices is the spectral broadening, which is due to the strong intermolecular interactions. As a consequence of this, the IR, Raman, or UV-VIS spectra are usually not sufficient for unambiguous identification of distinct molecules. Very often, only the functional groups present in the ice can be determined on the basis of characteristic vibrations; isomers possessing the same functional group or molecules of a homolog series are indistinguishable.[22] The mass distribution of the ionized desorbing gaseous molecules, recorded by a QMS, frequently does not provide adequate information to distinguish between isomers. Furthermore, due to fragmentation, the electron impact ionization is usually insufficient to identify elusive, weakly bonded molecules that might be abundant under certain conditions in space.[22]

To overcome the spectral resolution problem upon the investigation of astrophysical ices, the astrochemical species can be studied in inert (*e.g.*, noble gas) matrices, by so-called matrix isolation technique.[23,24] This method, of course, has serious limitations, since the environment is very different from that in astrophysical ices. Nevertheless, spectral information and some limited astrochemically relevant processes, *e.g.*, photodecomposition can be studied, at least as a first approximation by this method. Experiments using matrix isolation method, which aim to be the most useful for astrochemical studies, apply solid *para*-$H_2$ as a host. The main advantage of this host is that the intermolecular interactions are extremely small; therefore, the sharp spectral lines are suitable for dependable molecular identification. Further advantage of this method is that H atoms can efficiently be generated in this matrix, and the H atoms rapidly diffuse in there by quantum tunneling.[25–29] This provides excellent conditions to study low-temperature H atom addition and subtraction reactions, which are important processes in dark molecular clouds. The advantages and limitations of this method for astrochemical studies were discussed recently.[30] In addition, $H^+$ (in the form of astrochemically important $H_3^+$ and $H_n^+$ species) can also be effectively produced in solid para-$H_2$, providing a possibility to study protonation reactions and protonated astrophysical species.[31]

It is very challenging to complement the usual mass analysis of the desorbing species upon the TPD process with a spectroscopic method, because the concentration of these species in the gas phase is very low, and the spectra can be accumulated for a short period of time at a given desorption temperature. Due to this difficulty, conventional low-signal-to-noise and slow gas-phase spectroscopic methods cannot be applied for this purpose, and there is a limited number of astrophysical ice setups for which special designs are used to overcome this problem. Among these, Theulé et al. showed by a proof-of-principle study that high-resolution microwave spectra of the desorbing species can be obtained by a chirped-pulse Fourier-transform spectrometer if the gas pressure reaches ca. $6 \cdot 10^{-6}$ mbar.[32] A similar submillimeter/far-IR setup (SubLIME) was constructed by Weaver and coworkers.[33] Kaiser and coworkers demonstrated in a number of studies that photoionization (PI) reflectron time-of-fight mass spectrometry (ReTOF-MS) can be effectively coupled with an astrophysical ice setup to analyze the species desorbing during the TPD process.[11] On their Keck machine the intense monochromatic radiation is produced by third-harmonic generation or four-wave mixing of laser beams (ESI of ref. [34]). The main advantage of this setup is that the ionization potentials belonging to different mass channels can be determined by repeating the experiments using different photon energies. Comparison of the measured ionization potentials to literature data or quantum-chemical computations makes it possible to differentiate between different isomers. A further advantage of the PI-ReTOF-MS compared to the electron-impact QMS gas analyzer is that fragmentation can be considerably reduced by tuning the photon energy just above the ionization energy.[22]

In this paper we describe the construction, and demonstrate the strengths of a new setup: VIZSLA[35] [ˈviʒlɒ], (Versatile Ice Zigzag Sublimation setup for Laboratory Astrochemistry) that is primarily constructed to provide IR spectroscopic information on the desorbing species from the surface of astrophysical ices upon the TPD process. The setup comprises two cryostats. The ice is deposited onto and processed on a substrate mounted on the first cryostat. In a chosen temperature range of the TPD process, the desorbing molecules are mixed with an inert gas and deposited onto a substrate mounted on the second cryostat. Therefore, matrix-isolation IR (or UV-Vis) spectra of the species accumulated this way can be recorded. The molecules desorbing either from the first or the second cryostat can also be analyzed by a traditional way, using a QMS. VIZSLA is equipped with a laser OPO photon source (213–2800 nm), an H-atom lamp (Lyman-α, 121.6 nm), an electron gun, and an H atom beam source to trigger reactions. It also consists of a *normal*-H$_2$ to *para*-H$_2$ converter, providing the possibility for investigation of H atom reactions in solid *para*-H$_2$.

## II. SYSTEM DESCRIPTION

The instrument has been designed to fulfill the requirements for both solid-phase astrochemical ice experiments, and *para*-H$_2$ matrix isolation studies with the extra capability of sublimation and recondensation of the sample between two cold heads. (This process is referred to by the phrase of 'Zig-zag Sublimation' in the name of the setup.) In order to record relevant data on astrochemical ices, one needs UHV conditions, while for *para*-H$_2$ studies it is inevitable to reach temperatures below 3.5–3.8 K. To achieve an effective sublimation-recondensation process, the cold heads need to be close to each other. Furthermore, to avoid contamination of the secondary cryostat used for the recondensation, it has to be separated from the other, primary cryostat for the time when the ice is being deposited and processed on the substrate of the primary cryostat.

### A. The UHV chamber

The main component of the system is a cylindrical UHV chamber that has one coaxial interface port on the bottom, and one on the top lid to accept two independent cryostat units (Fig. 1). There are 14 ports on the side of the chamber in the same plane. One large port (DN160CF) is dedicated to the turbomolecular pump (Leybold TURBOVAC MAG W 600 iP) whereas three medium-sized ones (DN100CF) for the sample inlet on the front and for the two viewports on the sides. Furthermore, there are six smaller (DN40CF) flanges on the front and four on the backside of the chamber forming three plus two symmetric pairs, respectively. Those are used to connect different manipulation tools, irradiation sources, and various analytical tools to the chamber. Since the flanges are identical, there is a lot of freedom in the configuration of the available tools so the chamber can be adapted to the requirements of the specific experiments to increase the versatility of the system. The lid of the chamber has a differentially pumped seal, and it can be lifted up to give access to the inside components. Hence, everyday maintenance, *e.g.*, cleaning, does not require time-consuming disassembly of the system. The chamber is supported by a metal frame and surrounded by an optical table to give a solid base for the necessary optical arrangements around the chamber.

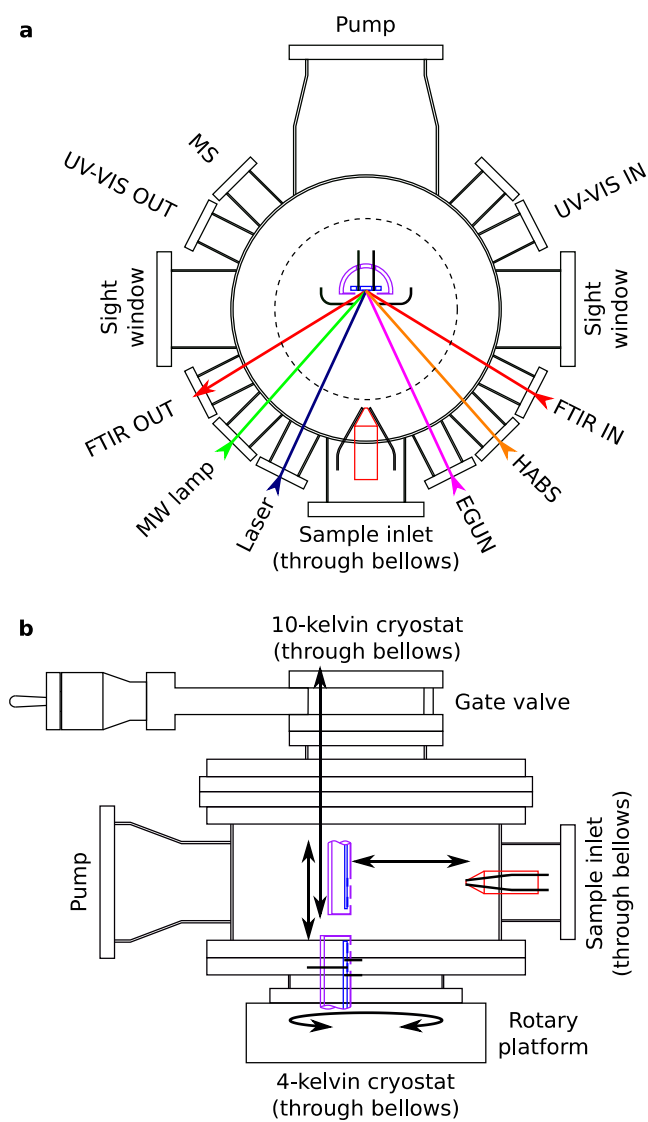

**FIG. 1.** The main UHV chamber of VIZSLA. **(a)** Horizontal section, **(b)** vertical section. FTIR IN and OUT: entrance and exit of the IR beam of the FT-IR spectrometer; UV-VIS IN and OUT: entrance and exit of the light beam of the UV-Visible spectrometer; HABS: hydrogen atom beam source; EGUN: electron gun; Laser: tunable OPO laser beam (213–2800 nm); MW lamp: microwave discharge H-atom lamp (H Lyman-α source); MS: quadruple mass spectrometer. Except for FTIR IN and OUT and UV-VIS IN and OUT, the other components can easily be exchanged on the ports according to the requirements of the experiment. (The attachment of the UV-Visible spectrometer to the systems is still under construction.)

When the 10-kelvin cryostat is in its upper position (see below), it can be separated from the main chamber by a gate valve. An auxiliary turbomolecular pump (Edwards Next85H CF63 NW16 80W) maintains the vacuum in the separated section while the gate valve is closed (Fig. 1).

The two turbomolecular pumps of the UHV chamber share the same fore vacuum system consisting of a roots pump (Leybold RUVAC WAU 251) built together with a two-stage rotary vane pump (Leybold TRIVAC D 40 B).

The pressure in the chamber is monitored by convection (Lesker KJLC 275 series) and EB-degas Bayard-Alpert Iridium Hot Cathode (Lesker G8130) ionization gauges. The chamber, excluding heat-sensitive parts, can be baked up to 80°C. The baking system of the UHV chamber is divided into 12 sections. Each section is equipped with a K-type thermocouple and an Autonics TX4S-B1R temperature controller. In addition, to prevent overheating of the cryostats, the temperature of the flanges connecting the cryostats and the UHV chamber and the temperature of cryostat heads are monitored during the baking. The heating elements are made of silicone isolated heating cables. For remote control purposes, the Autonics TX4S-B1R temperature controllers are connected via an RS485 local network and an Autonics SCM-WF48 communication converter to a PC running the DAQMaster program. After baking, the system can be evacuated to the $10^{-9}$ mbar region at room temperature.

**B. 4-kelvin cryostat unit**

The 4-kelvin cryostat unit (Sumitomo RDK-415D) is attached to the bottom of the chamber through a differentially pumped rotary platform and bellows. This allows the cryostat head to be translated vertically and rotated around the vertical axis by 360°. One of the end positions of the translation belongs to the sample substrate being in the center of the chamber. The cryostat is completely out of the middle section of the chamber at the other end position. An offset head made of OFHC copper is bolted to the second stage of the cryostat equipped with a heater, and a Si diode thermometer. Indium gaskets were used between mating surfaces to ensure good heat conductivity. An aluminum radiation shield bolted to the first stage of the cryostat achieves thermal isolation. The lowest available temperature is 3.1 K, which can be reached within a dynamic 1.5 h cool down time.

## C. 10-kelvin cryostat unit

The 10-kelvin cryostat unit (Sumitomo CH202) is attached to the top of the chamber through a gate valve and bellows, allowing the cryogenic head to be translated vertically. At one end position of the vertical travel, the sample substrate is in the center of the chamber, at the other end position the cryostat is fully retracted into the bellows section permitting the gate valve to be closed for complete separation of the 10-kelvin system from the main chamber. The structure of the head attached to the second stage, and the radiation shield attached to the first stage are similar to those of the 4-kelvin unit except the 10-kelvin head is slightly smaller, and has a *para*-$H_2$ converter attached to the cold head as well. This double use of the second cryostat made the system more compact and economical to build. The converter was inspired by an earlier design but slight modifications were required for adaptation to UHV conditions.[36] The $Fe_2O_3$ catalyst-filled 1/8" OD OFHC copper tube was coiled around an OFHC copper block previously wrapped in an indium foil. The coil was tightened by aluminum clamps. The catalyst block has a thermocouple, as well as a heating element built into it, and it can be dismantled from the cold section in order to allow thermal activation of the catalyst. Stainless steel capillaries carry the *normal*-$H_2$ gas into and the *para*-$H_2$ gas out of the converter through all-metal valves. Despite this small cold head is heavily overloaded, it can still reach temperatures below 12 K, but with the drawback of a long (~9 h) cooldown time. This turned out not to be a serious constraint if it is taken into consideration upon planning the experiments.

## D. Sample heads

A 10 mm × 10 mm polished reflective (presently gold) coated silver substrate is mounted on both the 4-kelvin and the 10-kelvin sample heads (Fig. 2). Indium gaskets were used to minimize the temperature difference between the substrate and the OFHC Cu block. Horizontally the full 10 mm size of the substrate is available, vertically only 8 mm can be exposed. The temperatures of the cold heads are monitored and controlled by a Lakeshore 365D temperature controller. There is one extra mounting point on both heads, those can be used either for mounting an alignment target or a standard Quartz Crystal Microbalance (QCM). Currently, the 4-kelvin cold head is equipped with a fluorescent ZnS target (Specs). There are cutout windows on the radiation shield in order to give access to the sample and the target (or QCM) surfaces.

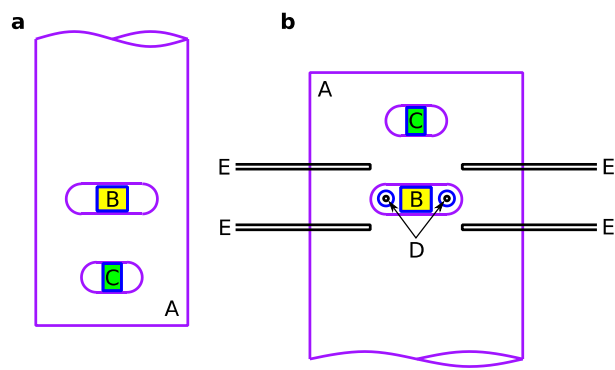

**FIG. 2.** The two cryogenic heads of VIZSLA. **(a)** The 10-kelvin cryostat head; **(b)** The 4-kelvin cryostat head. A: radiation shield; B: gold-coated silver substrate; C: fluorescence plate (electron irradiation alignment target); D: dual-capillary gas inlet; E: quadruple-capillary gas inlet.

The offset construction and the rotation of the 4-kelvin cryostat allow both heads to be present in the center of the chamber facing each other at the same time making possible the sublimation of the sample from one of the substrates and the recondensation on the other one. Minimal distance (~30 mm) had been chosen between the sample heads that still gives enough gap for the FT-IR beam to pass through without clipping, while the heads face each other. This maximizes the efficiency of the evaporation-recondensation process, while still enables IR spectroscopic measurements on the 10-kelvin cryostat without interrupting the TPD/redeposition process or moving the 4-kelvin head out of the IR beam path. Some possible geometrical arrangements accessible by the rotation of the 4-kelvin head and the translation of both cryostat heads are shown in Fig. 3.

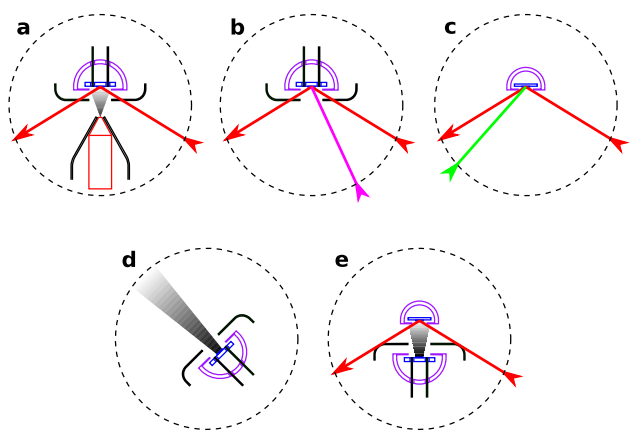

**FIG. 3.** Some possible geometrical arrangements of the cryostat heads. **(a)** Sample deposition onto the substrate of the 4-kelvin head with simultaneous FT-IR spectroscopic measurement. **(b)** Electron irradiation of the sample deposited onto the substrate of the 4-kelvin head with simultaneous FT-IR spectroscopic measurement. **(c)** H Lyman-α irradiation of the sample deposited onto the substrate of the 10-kelvin head with simultaneous FT-IR spectroscopic measurement. **(c)** Quadruple MS measurement during a TPD process of the sample on the substrate of the 4-kelvin head. **(d)** Redeposition, *i.e.*, deposition of the mixture Ar gas (entered through either the dual- or the quadruple-capillary gas inlet) and the molecules desorbed during a TPD process of the sample on the substrate of the 4-kelvin head onto the substrate of the 10-kelvin head with simultaneous FT-IR spectroscopic measurement.

## E. Sample inlets

The system has a main sample inlet on the front side, equipped with a stainless steel Knudsen evaporation/decomposition cell rated up to 350°C, and two dual-capillary gas sample inlets, allowing simultaneous deposition of one non-volatile sample, and two gas-phase mixtures (Fig. 1.). Flow rates are regulated by leak valves (Lesker Ultra-fine, VZLVM940R). The gas jets only meet at the front of the cold surface, so one can also deposit such kinds of gases that would otherwise react with each other. The sample inlet connects to the system through bellows, so it can be positioned closer or farther away from the substrate, and can also be retracted to clear the center of the chamber to make space for the rotation of the 4-kelvin cryostat head.

Besides the main sample inlet used for initial deposition, the 4-kelvin cryostat unit also has two independent secondary gas inlet capillary systems supported by room-temperature stands around the cold head (Fig. 2). A dual-capillary inlet gets through holes drilled on the cold head without touching the cold surfaces, and a quadruple-capillary inlet located completely outside the radiation shield surrounds the sample. For sublimation-recondensation experiments, the gas sample inlets, located on the 4-kelvin head, give the opportunity to co-deposit the molecules entered from the primary cryostat to the gas phase with argon (or another) gas by flooding the space between the heads in order to form a matrix-isolated sample. These inlets can also be used for introducing other reactive components. The 10-kelvin system has no gas sample inlet but there is space and mounting stand for one quadruple-capillary version so it can be implemented later if it is necessary.

For highly corrosive volatile substances not compatible with stainless steel or for medium volatility compounds sublimating under high-vacuum conditions at room temperature an extra PTFE tube inlet is also available, and it can be mounted on one of the DN40CF flanges. (The construction and building materials of that inlet compromises the vacuum level. Due to its fixed position, depending on its configuration this inlet might also interfere with the rotation of the 4-kelvin cryostat. Because of these reasons, if it is not mandatory for an experiment, the application of this unit should be avoided.) Future plans include designing a pyrolysis source that could substitute the Knudsen cell if needed. For samples with medium volatility, a cooling system for the Knudsen cell inlet is also under construction.

## F. Gas manipulating system of VIZSLA

Gas and volatile samples, and the *para*-H$_2$ are manipulated with a metal gas mixing vacuum-line equipped with valves, glass bulbs, metal cylinders, and various (piezo, Pirani, capacitance, and Penning) pressure gauges (Fig. 4). The vacuum-line can be evacuated to a high-vacuum level (in the range of 10$^{-7}$ mbar) and baked to help the desorption of the contaminants from the walls. Its 2 baking sections are part of the baking system described in section II.A.

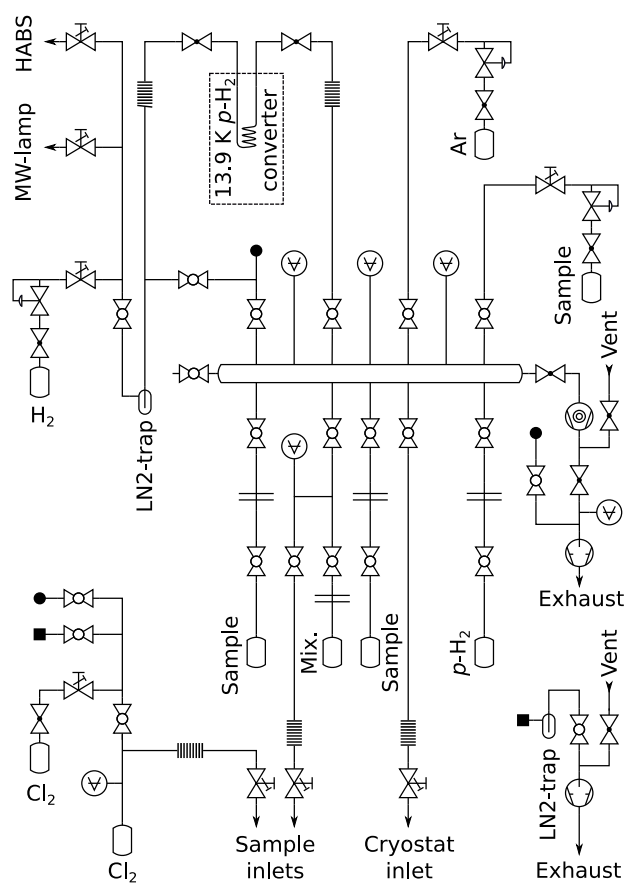

**FIG. 4.** The gas manipulating system of VIZSLA.

The vacuum line for sample manipulation has its own vacuum system consisting of a turbomolecular pump (Edwards Next85H NW40 NW16 80W) and a rotary vane pump. For evacuating corrosive and hazardous gases an alternative rotary vane pump equipped with a liquid $N_2$ cooled trap is used to protect the high-vacuum system and the operators.

### G. Irradiation and particle sources

Various light and particle sources are available for the experiments. The sample can be irradiated through a quartz window with a tunable optical parametric oscillator laser source (OPO, GWU versaScan-ULD & GWU uvScan) pumped by the third harmonic of a nanosecond Nd:Yag laser (Spectra Physics LAB150-10). The laser system is tunable from the UV to the NIR (213–2800 nm).

For mimicking the interstellar H Lyman-$\alpha$ radiation a microwave discharge H-atom lamp can be used. The discharge tube of the lamp is made of quartz, designed according to Ligterink *et al.*, with two ports on the side (F-shaped) and can be mounted onto a differentially pumped LiF window on the chamber.[37] The lamp can be powered by an EMS Microtron 200 Mk3 microwave generator ($P = 100$ W). $H_2$ gas is entered into the quartz tube through the inlet closer to the window, while the other port of the lamp tube was connected to a rotary vane pump. The pressure is measured between the pump and the outlet of the lamp and for usual operation, it is set to 0.4 mbar. (It was shown by Ligterink *et al.* that for this applied geometry, this pressure is optimal to get the highest intensity of the astrophysically important H Lyman-$\alpha$ radiation compared to the broad-band UV background.) Naturally, any standard lamp (like Hg-Xe discharge lamp or LED sources) can be used as well.

Available particle sources include a commercial electron gun (Specs EQ 22) and a thermal hydrogen atom beam source (MBE Komponenten HABS 40) equipped with a bent quartz tube to cool down the thermally generated H atoms.

### H. Analytical tools

The most exterior DN40CF flanges on the front side of the chamber are equipped with ZnSe windows (Lesker, 2−3/4"UHV VPZL-275UZ) and dedicated to the main analytical tool of the system, FT-IR spectroscopy. The beam of the spectrometer (Bruker Invenio-R) is reflected into the chamber with mirror optics on the left side. The IR beam hits the surface of the film at about 60° angle of incidence (AOI), passes through the sample, then is reflected from the substrate, and passes through the film again. The doubled and increased beam path improves the sensitivity of the system by a factor of 4 in the case of *para*-$H_2$ matrix or at least by a factor of 2 in the case of films with a high refractive index, compared to a normal incidence transmission geometry. The beam of the spectrometer leaves the UHV chamber on the opposite side and it is detected by a liquid $N_2$ cooled external MCT detector (Bruker, mid-band) of the spectrometer. The sample on either the 4-kelvin or the 10-kelvin cryostat can be measured with the same spectrometer separately by moving one or the other cryostat into the beam path. The smallest possible distance was chosen between the sample heads that still gives barely enough gap for the FTIR beam to pass through when the heads are facing each other. This maximizes the efficiency of the sublimation-recondensation process while it still enables the continuous monitoring of the sample on the secondary cryostat during the deposition. Currently, the spectral cutoff at the low-wavenumber side is limited by the ZnSe windows (600 cm$^{-1}$). Differentially pumped wedged KBr windows are under construction to increase the available spectral window.

The implementation of UV-VIS spectroscopic detection (Ocean Optics HR2000) is under construction. Since for this application the DN40CF flanges on the backside of the chamber can be used, it will be available only for the 4-kelvin system, as its cryostat can be rotated into the required position. Alternatively, the Bruker Invenio-R FT-IR spectrometer can be upgraded in the future with its commercially available UV extension, which could provide a convenient UV-VIS spectroscopic study of the films or matrices deposited onto the substrate of either of the 4-kelvin or the 10-kelvin cryostats.

The system will also be equipped with an optical film thickness monitoring system based on He-Ne laser interference.

For the detection of species in the gas phase, the chamber is equipped with a QMS (Pfeiffer Prisma QMA 200).

## III. EXPERIMENTAL PERFORMANCE

### A. *Para*-$H_2$ conversion

Matrix isolation techniques have long been used in spectroscopy to assess species in a nearly interaction-free environment. In this kind of experiments, the sample molecules (guests) are co-deposited with an inert gas called host (Ar, Ne, Kr, Xe, $N_2$) in great excess onto a surface cooled to a low temperature, usually between 10–30 K depending on the gas used.[24,38] Consequently, the molecules are frozen into a solid matrix, isolated from each other by the host atoms/molecules. The technique requires a high vacuum to achieve the necessary thermal isolation and to keep away impurities from the atmosphere. The gases listed above are used most frequently but more exotic matrix materials could also be used; for instance reactive compounds such as CO or *para*-$H_2$, which latter has some unique properties making it a perfect candidate to perform matrix-isolation experiments. First, due to the extremely weak interaction with the guest molecules, the bandwidth of the IR absorption peaks in this solid quantum material is even smaller than in the more conventional matrix hosts listed above. Secondly, since H atoms can effectively be generated in the matrix owing to the quantum diffusion effect that efficiently speeds up the H atom movement within the matrix,[26–29] H atom reactions can conveniently be followed in *para*-$H_2$. It should be noted, however, that it has its limitations when it comes to the direct comparison with astrophysical ices or ice mixtures.[30]

$H_2$ has nuclear spin isomers, the *ortho*- and *para*-$H_2$, with the latter being thermodynamically slightly more stable. Due to the near degenerate energy of the two forms, the *ortho* to *para* ratio is 3 : 1 at room temperature. Upon lowering the temperature, the ratio of the lower energy *para* form is increasing, but due to kinetic control, the equilibrium can be reached very slowly. The rate of the conversion process can be increased by a paramagnetic catalyst, which is $Fe_2O_3$ in our case. Before entering the converter, impurities of the $H_2$ gas (>99.999%, Messer) were trapped inside a copper coil cooled by a liquid $N_2$ bath. The conversion temperature was selected to be 13.9 K, which was found to be the lowest for our converter, to achieve continuous flow (approximately 22 sccm $min^{-1}$) without condensation. The freshly generated *para*-$H_2$ is collected in a glass bulb and can be used for the experiments thereafter, for instance for mixing with the sample material or for directly introducing it to the main chamber using one of the capillary arrays. To achieve better purity, the first 2×~500 mbar·L of *para*-$H_2$ is thrown away. At the applied 13.9 K conversion temperature, the *ortho* form has a mixing ratio of approximately 50 ppm.[25] Since impurities can catalyze the *para* to *ortho* back conversion, the *ortho* impurity has to be checked for in each experiment. The relative amount of *ortho*-$H_2$ impurity can be easily estimated based on the NIR spectrum taken from the deposited matrix. According to Tam and Fajardo, the absence of the $Q_1(0) + S_0(1)$ transition belonging to *ortho*-

H$_2$ at 4740 cm$^{-1}$ indicates an *ortho*-H$_2$ concentration less than 200 ppm, whereas the concentration is expected to be around 20000 ppm if its intensity is comparable to the neighboring S$_1$(0) + S$_0$(0) transition of *para*-H$_2$.[39] More on the two forms of H$_2$ and on the nuclear spin conversion can be found in the comprehensive review written by Tsuge and Lee.[25] In order to demonstrate the purity of *para*-H$_2$ obtained by our setup, Fig. 5 displays the NIR spectra of *normal*-H$_2$ (3 : 1 *ortho*-H$_2$ : *para*-H$_2$) and *para*-H$_2$ deposited onto the substrate of the 4-kelvin cryostat at 3.1 K. To estimate numerically the *ortho*-H$_2$ impurity, the relative integrated absorptions of the bands of the spin isomers were used. In the *normal*-H$_2$ ice, the *ortho*-H$_2$ band at 4740 cm$^{-1}$ has an integrated area of 1.722 cm$^{-1}$, whereas integrating that of the *para* form between 4520–4495 cm$^{-1}$ yields 0.918 cm$^{-1}$, and a ratio of 1.88 for the two bands. The respective band areas in *para*-H$_2$ were found to be 0.0018 cm$^{-1}$ and 5.855 cm$^{-1}$, giving a ratio of 0.000307. Taking these ratios, and keeping in mind that the *ortho*-to-*para* ratio in *normal*-H$_2$ is 3 : 1, an *ortho*-to-*para* ratio of 0.000490 (490 ppm) can be evaluated, which is typical for our *para*-H$_2$ matrix experiments.

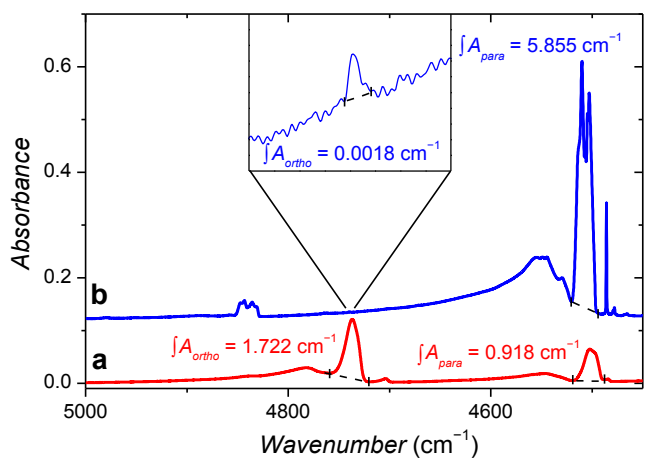

**FIG. 5.** H$_2$ combination bands in the NIR spectra. (**a**) *normal*-H$_2$ (3 : 1 *ortho*-H$_2$ : *para*-H$_2$) after introducing 105.9 mbar·L *normal*-H$_2$ into the main chamber over a deposition time of 108.0 min (meaning an average flow rate of 0.98 mbar·L min$^{-1}$); (**b**) *para*-H$_2$ obtained by our converter that was deposited onto the substrate of the 4-kelvin cryostat at 3.1 K after introducing 184.0 mbar·L *para*-H$_2$ into the main chamber over a deposition time of 167.5 min (average flow rate was 1.10 mbar·L min$^{-1}$). The spectra are offset for clarity. The inset shows the enlarged spectral region around the 4740 cm$^{-1}$ band due to *ortho*-H$_2$ in our deposited *para*-H$_2$ matrix. The integrated band areas of the respective bands are visualized on the figure. The black vertical lines show the integration boundaries, whereas the dashed black lines highlight the baselines used when integrating the bands.

### B. H atom generation and H atom reactions in solid *para*-H₂

The method of generating H atoms in a *para*-$H_2$ matrix was originally established by the groups of Anderson and Lee during which numerous H addition, as well as H abstraction reactions, have been described.[28,29,40–55] According to this procedure, the studied compound is deposited in a solid *para*-$H_2$ matrix that contains $Cl_2$ in a ratio of ca. 1 : 500–1 : 2000. Cl atoms can be generated from $Cl_2$ by photodecomposition, in the range of ca. 250–400 nm. A wavelength for a given experiment should be selected, which does not photolyze the molecule under investigation. Practically we use our OPO system for this photolysis. In contrast to noble gas matrices, due to the diminished cage effect in solid *para*-$H_2$, the Cl atoms do not recombine.[25] The forming Cl atoms can be monitored by checking for the spin−orbit (SO) + $Q_1(0)$ transition at 5095 $cm^{-1}$ that is due to the simultaneous excitation of a Cl atom and a vibrational ($v' = 1$, $J' = 0 \leftarrow v'' = 0$, $J'' = 0$) transition of a *para*-$H_2$ molecule (Fig. 6). Although *para*-$H_2$ molecules do not react with Cl atoms, they can be forced to do so by exciting the $Q_1(1) + S_0(0)$ and $Q_1(0) + S_0(0)$ combination vibration of *para*-$H_2$ by irradiating the matrix with either with a broad-band NIR source (*e.g.*, that of the FT-IR spectrometer) or 2217 nm laser radiation (generated by our OPO system):

$$Cl + H_2 \rightarrow HCl + H$$

The broad-band NIR source has the disadvantage, that in our case it has lower intensity, furthermore, it can induce the H atom abstraction or H atom addition reactions by vibrationally exiting the molecule in study. This usually unwanted IR induction can be avoided by using the 2217 nm laser radiation, however, the high energy pumped into the matrix can increase the diffusion rate of the H atoms and can increase the average kinetic energy of the H atoms. These effects should be studied in details in a further study.

The formation of H atoms upon the NIR irradiation can be monitored by measuring the absorption peak of byproduct HCl molecules at 2895 $cm^{-1}$ in the MIR spectrum. FT-IR spectra in the MIR range can be recorded in real time without switching off the NIR irradiation source using a low-pass filter with a cutoff wavelength of 3960 $cm^{-1}$ in front of the detector (Fig. 6).

There are alternative ways to produce H atoms, such as by the (193–380 nm) UV photolysis of $H_2O_2$ present in the matrix.[56] Accordingly, $H_2O_2$ yields two OH radicals upon dissociating, which spontaneously produce H atoms in the immediate reaction with $H_2$.[27,41] The combination of these different procedures helps discriminate the products with H atoms from the possible minor side products generated in a reaction with Cl atom.

Examples for H atom reaction in solid *para*-H₂ conducted on VIZSLA include the successive hydrogenation of SO and SO₂, and the H atom abstraction from glycine resulting in C$_\alpha$-glycyl radical.[57,58]

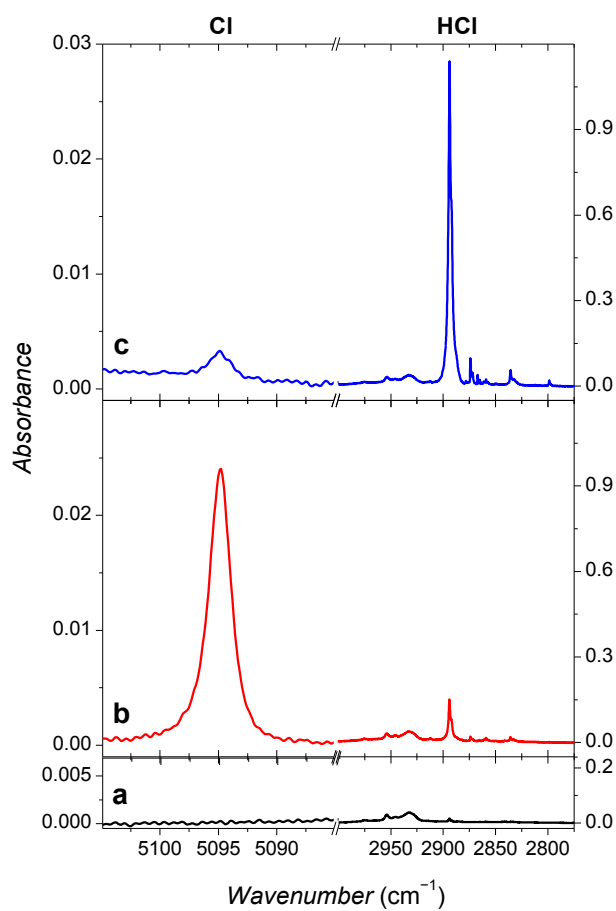

**FIG. 6.** IR spectra showing the change of Cl and HCl ratio in a *para*-H₂ matrix during H atom generation. The bands at 5095 and 2895 cm⁻¹ correspond to Cl and HCl, respectively. (**a**) Spectrum recorded after deposition of the *para*-H₂ : Cl₂ (∼800 : 1) sample on the substrate of the 4-kelvin cryostat. (**b**) Spectrum recorded after irradiating the matrix with a UV laser light of 365 nm for 90 min ($P \approx 1.8$ mJ pulse energy @ 10 Hz at the laser aperture, 100 µJ pulse energy at the substrate). (**c**) Spectrum measured after subsequent irradiation with a NIR light of 2217 nm for 44 min ($P \approx 3.4$ mJ pulse energy @ 10 Hz at the laser aperture, 0.57 mJ pulse energy at the substrate). The laser light was made divergent using a lens in order to enlarge the laser beam area so that the whole substrate could be uniformly irradiated.

## C. Generation of protonated species in solid *para*-H₂

*Para*-H₂ ice is also suitable for the investigation of protonated species. An electron gun can be used to generate H⁺ (and H$_n$⁺) ions. In practice 200 eV to 2 MeV electrons with a current from 0.1 to 60 µA are used.[31,59] With our setup 2.0 to 5.0 keV electrons with currents of 20 to 500 nA were tested. We have found that electron irradiation can be applied during deposition but to avoid interaction of the electrons with the gold surface, which was found to cause deterioration of the matrix, initially a *para*-H$_2$ layer without electron irradiation has to be deposited. The minimum thickness of this layer was estimated by the CASINO program[60] and was found to be 650 and 3000 nm for 2.0 and 5.0 keV electrons, respectively. To avoid Coulomb explosion, the electron gun was operated periodically, using 5–15 min on/off cycles, while continuously depositing the matrix. To show the performance of our setup, the experiment of Tsuge *et al.* for the generation of XeHXe$^+$ ions[59] were reproduced and the spectrum taken during this experiment is displayed in Fig. 7.

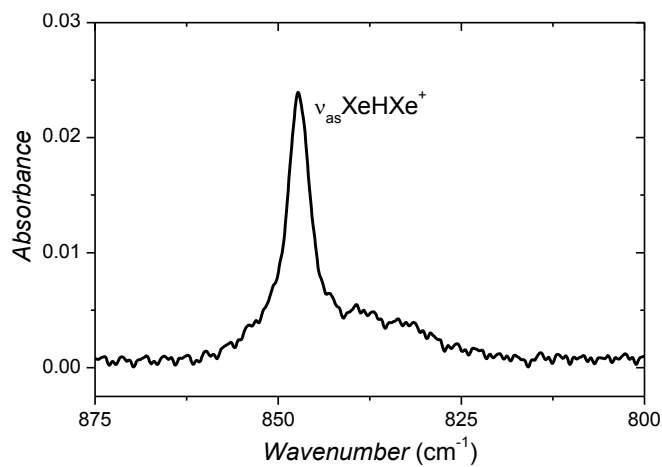

**FIG. 7.** MIR spectrum of XeHXe$^+$. The band at 847.3 cm$^{-1}$ appeared upon irradiating the *para*-H$_2$ matrix containing 1% Xe with 5 keV electrons. The electron emission current was kept at 30 nA. The deposition time was 194 minutes.

### D. TPD – matrix-isolation redeposition experiments: Demonstration of feasibility

As mentioned in the Introduction, the other function of VIZSLA allows the processing of astrophysical ices followed by sublimation (*i.e.*, temperature-programmed desorption, TPD) and parallel matrix-isolation redeposition for IR spectroscopic measurement. Since the molecules isolated in an inert, weakly-interacting (*e.g.*, Ar) matrix have sharp bands and well-resolved IR spectra, by comparing the measured spectra of the redeposited species with literature matrix-isolation IR spectra (or with computed IR spectra) reliable, unambiguous molecular identification can be made. First, we demonstrate the feasibility and the strengths of the method, then we provide efficiency measurements of the redeposition, finally, we show two applications on a processed astrophysical ice.

In order to demonstrate the feasibility of the method, we have studied three different ices: sulfur dioxide ($SO_2$), methanol ($CH_3OH$), and the 1 : 1 : 1 mixture of methanol ($CH_3OH$), ethanol ($CH_3CH_2OH$), and isopropanol (($CH_3$)$_2$HOH). The first two systems were chosen because of the possible formation of clusters at the sublimation with different extent due to different strengths of secondary chemical bonds. This effect has to be explored in details, because the possible deposition of different clusters could result in a spectral broadening, and therefore, could seriously limit the application of the method. The mixture was chosen to demonstrate and test the analytical strength of the method for discriminating between molecules possessing the same functional group.

In each case, the gas (or gas mixture) was first deposited on the substrate of the 4-kelvin cryostat unit at 3.1 K. Then a TPD process was started. Some 10–20 K below the sublimation temperature of the sample the 4-kelvin was rotated and the 10-kelvin cryostat was lowered into the IR beam position, facing with the substrate of the 4-kelvin head. The Ar was introduced into the chamber, near the substrate by the dual-capillary inlet. The sublimed molecules mixed with Ar, and were deposited on the substrate of the 10-K cryostat at 11 K. During the matrix-isolation redeposition 64-scan averaged IR spectra were recorded each minute.

Fig. 8(a) shows the spectrum of an $SO_2$ neat film freshly deposited on the 4-kelvin substrate at 3.1 K. As expected, the spectrum of the film exhibits broad features due to the various intermolecular interactions within the solid. Fig. 8(b) shows the IR spectrum taken from the sublimed molecules redeposited in an Ar matrix on the substrate of the 10-kelvin cryostat. Based on the comparison of the two spectra one can immediately see the advantage of the higher resolution for analytical identification purposes. Although the lines (at 1351.1 and 1355.3 cm$^{-1}$, as well as at 1147.1 and 1152.2 cm$^{-1}$) are split due to different sites in Ar matrix, this is well-documented in the literature.[61] In this case, only a small amount of dimer was observed (at 1341.5 cm$^{-1}$) in the sample redeposited in Ar matrix.[62]

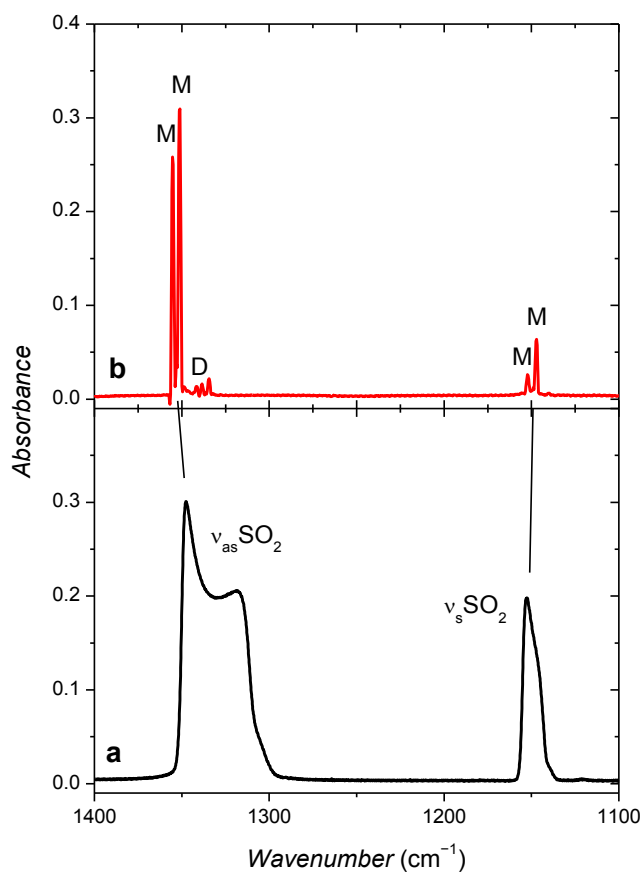

**FIG. 8.** MIR spectra of $SO_2$ in the 1100–1400 $cm^{-1}$ region. (**a**) $SO_2$ thin film freshly deposited on the substrate of the 4-kelvin cryostat at 3.1 K. (**b**) Sample redeposited and isolated in an Ar matrix on the substrate of the 10-kelvin cryostat at 11 K during a TPD process performed on the 4-kelvin cryostat between 85 and 115 K at a rate of 1 K min$^{-1}$. 'M' denotes the monomers, whereas 'D' stands for dimers.

The next example, in Fig. 9, demonstrates a similar experiment on $CH_3OH$. The same applies here as for $SO_2$, the IR absorption features of the $CH_3OH$ ice are broad and unresolved, which is greatly improved by redepositing the $CH_3OH$ and isolating it in an Ar matrix during the TPD process. Although $CH_3OH$ can form clusters by strong hydrogen bonds, and likely these clusters are present in the gas phase near the surface of the ice, almost exclusively the IR band of the monomer can be detected at 1033.5 $cm^{-1}$.[63] The dimers' and oligomers' weak peaks show up at 1029.0, 1038.4, and 1053.4 $cm^{-1}$, respectively.[64] The possible plausible explanation for this phenomenon is that the $CH_3OH$ aggregates are mostly destroyed in the gas phase due to the collision with Ar atoms, which are present in great excess in the space between the two substrates. Consequently, this experiment proves that the method also allows for the study of the monomeric form of substances known for their tendency to aggregate.

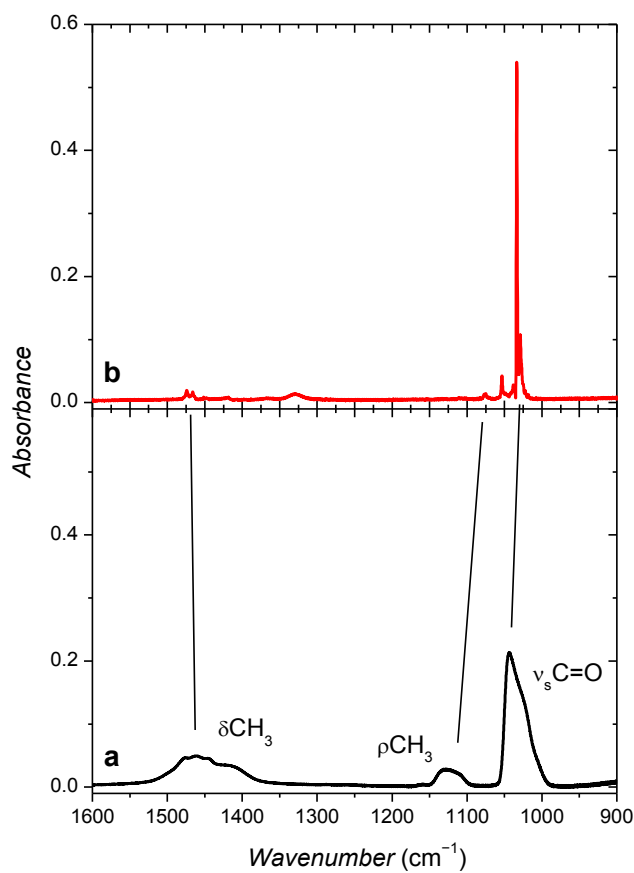

**FIG. 9.** MIR spectra of $CH_3OH$ in the 900–1600 $cm^{-1}$ region. (**a**) Thin film freshly deposited on the substrate of the 4-kelvin cryostat at 3.1 K. (**b**) Sample redeposited and isolated in an Ar matrix on the substrate of the 10-kelvin cryostat at 11 K during a TPD process performed on the 4-kelvin cryostat between 85 and 180 K at a rate of 1 K $min^{-1}$.

The third example (Fig. 10) demonstrates that the method is also suitable to study mixtures. The 1:1:1 mixture of methanol ($CH_3OH$), ethanol ($CH_3CH_2OH$), and isopropanol (($CH_3$)$_2$HOH) was processed similarly as the pure $SO_2$ and $CH_3OH$ ices. The lower panel depicts the spectrum of the mixture deposited on the substrate of the 4-kelvin cryostat at 3.1 K (Fig. 10(a)). On the basis of this spectrum, one might deduce that alcohols are present in the sample, but further analytical identification could be barely made on the safe ground. The upper panel shows three spectra recorded from the redeposited sample in an Ar matrix at three different stages of the TPD process. The sharp spectral lines allow the unambiguous identification of the compounds: $CH_3OH$ at 1033.6, 2847.6, 2955.6, 3005.5, and 3666.0 $cm^{-1}$,[64] $CH_3CH_2OH$ at 886.4, 1025.0,

1091.6, 1239.5, 1445.0, and 1463.1 cm$^{-1}$,[65] and (CH$_3$)$_2$CHOH at 813.7, 948.7, 1165.2, 1380.4, 1460.9, 2929.6, 2937.6, 2987.2, and 3639.5 cm$^{-1}$ (Supplementary Information of ref. [66]).

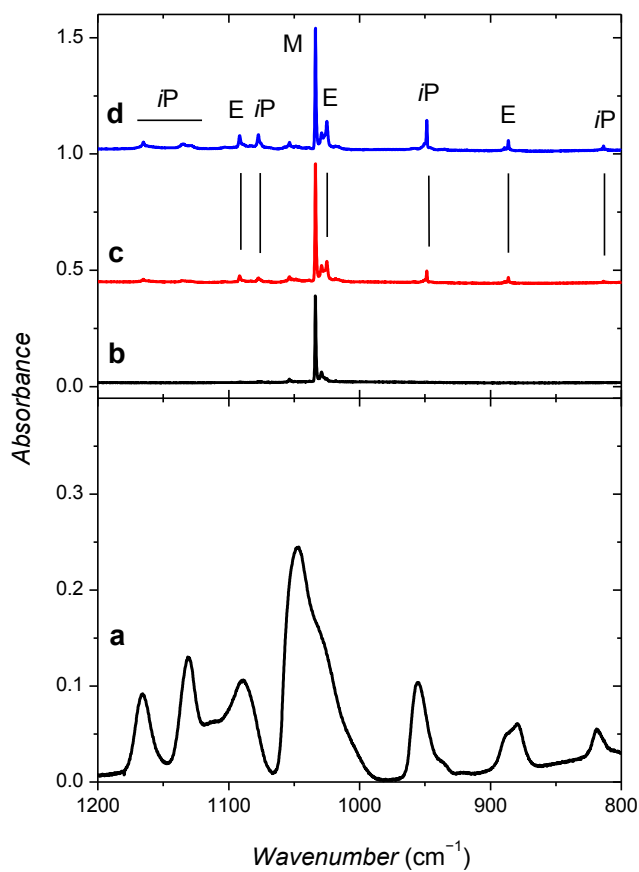

**FIG. 10.** MIR spectra of a methanol : ethanol : isopropanol (1 : 1 :1) mixture in the 800–1200 cm$^{-1}$ region. **(a)** Freshly deposited ice on the substrate of the 4-kelvin cryostat at 3.1 K. The spectral resolution was set to 4 cm$^{-1}$. **(b)-(c)** Sample redeposited and isolated in an Ar matrix on the substrate of the 10-kelvin cryostat at 11 K during a TPD process performed on the 4-kelvin cryostat between 85 and 180 K at a rate of 1 K min$^{-1}$. Spectra were obtained from the accumulated matrix-isolated sample on the 10-kelvin cryostat was recorded when the temperature of the substrate of the 4-kelvin cryostat was **(b)** 152 K; **(c)** 166 K; **(d)** 179 K. The resolution was set to 1 cm$^{-1}$. 'M', 'E', and '*i*P' denote the bands of methanol, ethanol, and isopropanol, respectively.

It is important to note and explore if the Ar flow shifts the sublimation temperatures. Typically, for isolation of molecules, Ar has to be used in an excess of 1 : 1000 or more. This means that it has a significant pressure at the surface. Although Ar is not reactive, the increased background pressure due to Ar might trap reactive intermediates on the surface, which species could sublime into the gas phase at normal (UHV) conditions. Then these species might react with each other on the surface resulting in secondary products. It cannot be ruled out, however, that this effect is counteracted, because high-temperature Ar atoms are bombarding the surface, and they possibly release molecules from the ice below the sublimation temperature. These should be investigated further in future studies when desorption of reactive species/radicals can be expected. To understand the effect of Ar flow the mass spectra of the subliming molecules and the IR spectra of the ice during a TPD process once without and once with Ar flow could be measured and compared.

### E. TPD – matrix-isolation redeposition experiments: Efficiency measurements

For future applications, it is important to quantify the efficiency of the sublimation – matrix isolation redeposition. The integrated absorption area of a band measured in a thin film cannot be compared directly to the one measured in an Ar matrix, because the intermolecular interactions might considerably affect the molar absorptions. Therefore, another experiment had to be carried out to find a definite answer for this question. In this case, first a sample : Ar = 1 : 1000 (in the case of $SO_2$) or 1 : 10000 (for $CH_3OH$) mixture was deposited on the substrate of the 4-kelvin cryostat at 3.1 K. Then, the sample was slowly heated to the sublimation temperature of Ar. After the complete sublimation of Ar (followed by the pressure gauge), a TPD – matrix-isolation redeposition experiment as described in Section III.E. was carried out.

Using this procedure, the integrated band areas measured in an Ar matrix on the 4-kelvin and on the 10-kelvin cryostat can safely be compared. However, it has to be kept in mind, that the determination of the efficiency by this way gives a lower estimate for the efficiency, because Ar is applied in great excess and during the sublimation of Ar some of the sample molecules can be lost.

Using the doublet IR band of the matrix-isolated $SO_2$ at 1351.1 and 1355.3 $cm^{-1}$ the integrated band areas of the freshly deposited matrix on the 4-kelvin cryostat and the one after TPD – redeposition onto the 10-kelvin cryostat were found to be 1.293 and 0.104 $cm^{-1}$, respectively (Fig. 11). This means that approximately 8.0% of the initial $SO_2$ molecules deposited onto the substrate of the 4-kelvin cryostat could be recaptured on the substrate of the 10-kelvin cryostat.

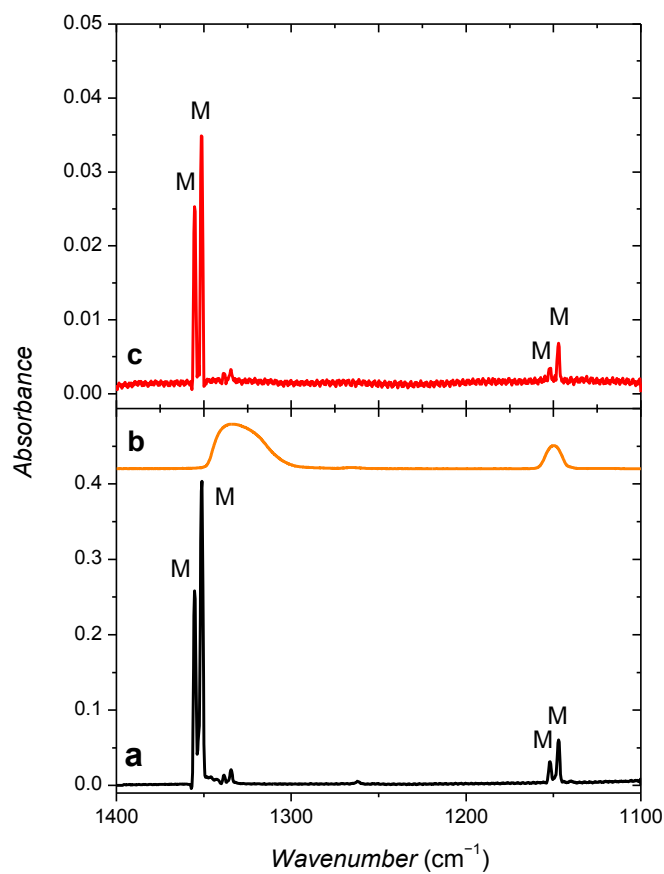

**FIG. 11.** MIR spectra of $SO_2$ in the 1100–1400 cm$^{-1}$ region. (**a**) Freshly deposited sample isolated in an Ar matrix on the substrate of the 4-kelvin cryostat at 3.1 K. The spectral resolution was set to 1 cm$^{-1}$. (**b**) $SO_2$ film after subliming the Ar at 70 K (spectral resolution: 4 cm$^{-1}$, offset for clarity). (**c**) Sample redeposited and isolated in an Ar matrix on the substrate of the 10-kelvin cryostat at 11 K after during a TPD process performed on the 4-kelvin cryostat between 85 and 180 K at a rate of 1 K min$^{-1}$. The spectral resolution was set to 1 cm$^{-1}$.

The efficiency of the redeposition process has been determined in a similar method for $CH_3OH$, too. Fig. 12 shows the IR spectra of the compound isolated in Ar matrix before and after the TPD process on the substrates of the 4-kelvin and the 10-kelvin cryostat, respectively. The integrated band areas of the 1033.8 cm$^{-1}$ feature are 2.547 and 0.251 cm$^{-1}$, respectively, meaning an overall redeposition efficiency of roughly 9.9%. This is in an agreement with the efficiency value determined for the $SO_2$.

The effect of the Ar flow rate on the recondensation efficiency was also studied. Three different Ar inlet rates were used (0.2, 0.7, and 2.0 mbar·L min$^{-1}$) but no significant differences could be found in the efficiency (0.132, 0.129, and 0.116, respectively). In contrast to this, the relative efficiency could be increased by some 50% to 0.192 when the quadruple-capillary array was used instead of the dual one (Ar flow was kept at 0.35 mbar·L min$^{-1}$). It should also be noted that when using lower Ar inlet rates, not only the CH$_3$OH monomer could be observed but also minor bands caused by the oligomers, which were not taken into account by this estimation discussed above.

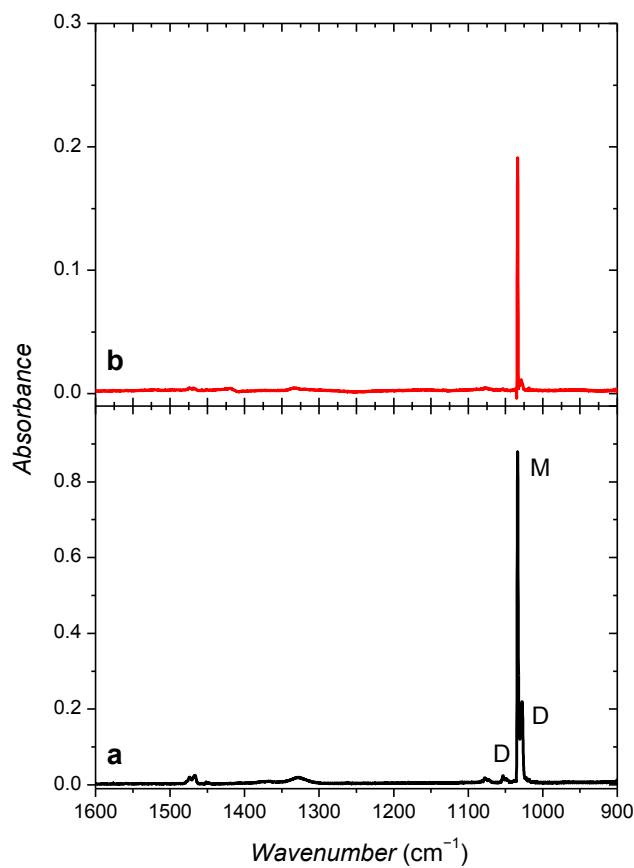

**FIG. 12.** MIR spectrum of CH$_3$OH in the 1100–1400 cm$^{-1}$ region (**a**) Freshly deposited sample isolated in an Ar matrix on the substrate of the 4-kelvin cryostat at 3.1 K. The spectral resolution was set to 1 cm$^{-1}$. (**b**) MIR spectrum of CH$_3$OH redeposited and isolated in an Ar matrix on the substrate of the 10-kelvin cryostat at 11 K after during a TPD process performed on the 4-kelvin cryostat between 85 and 180 K at a rate of 1 K min$^{-1}$. The spectral resolution was set to 1 cm$^{-1}$.

In conclusion, the efficiency of the TPD – matrix-isolation redeposition is roughly between 8 and 20%, depending on the sample, on the Ar flow rate, on the capillary used, and although, it is not tested, certainly on the distance of the two substrates. However, it should be noted, as demonstrated in Section III.E., due to the sharp lines in an Ar matrix compared to the ones observed in a neat ice, some bands can appear with larger absorbance in the redeposited matrix, even if 80–90% of the sample is lost during the redeposition.

### F. TPD – matrix-isolation redeposition experiments: Tests on processed astrophysical analogue ices

In the case of processed astrophysical analogue ices usually several products are generated upon the irradiation. These products have a mixing ratio usually much smaller than the compounds in the deposited, unprocessed ice; furthermore, the products are usually generated with very different efficiencies. In order to test the applicability of the TPD – matrix-isolation redeposition experiments for realistic processed ices, we have chosen two different systems, which were recently studied extensively.

The first test system is the irradiated methanol ice. Both the effects of the electron and the H Lyman-$\alpha$ irradiations have already been investigated.[67–69] The products were analyzed on the basis of the IR spectrum of the processed ice and on the basis of the mass spectra of the subliming molecules upon the TPD process. In the case of the two different irradiations, the same main products were identified. Here we have chosen to reproduce the experiment of the H Lyman-$\alpha$ irradiation of the methanol ice at 12 K. Since the TPD – matrix-isolation redeposition FT-IR analysis is expected to be less sensitive, to accumulate more products, a thicker film was deposited (absorbance of the most intense methanol band at ~1020 cm$^{-1}$ was ~0.9 vs 0.2) and a longer irradiation time (758 vs 60 min) was applied. (The longer irradiation time was also necessary, because in our setup the H lamp is farther from the substrate than in Ref[69].)

Fig 13. shows the IR spectrum of the deposited ice, the irradiated ice and spectra taken from the sample redeposited onto the substrate of the 10-kelvin cryostat. The different spectra of the redeposited sample collected at different temperature regions of the TPD process are shown. The products identified in the Ar matrix are collected in Table 1. As it can be seen most of the products identified by the earlier studies: methane ($CH_4$), carbon monoxide (CO), carbon dioxide ($CO_2$), formaldehyde ($H_2CO$), acetaldehyde ($CH_3CHO$), dimethyl ether ($H_3COCH_3$), methyl formate ($HCOOCH_3$), ethylene glycol ($HO(CH_2)_2OH$), and tentatively, ketene ($H_2CCO$), and ethylene oxide ($(CH_2)_2O$) are also trapped and identifiable in the spectrum of the matrix-isolated, redeposited sample. In many cases of the former studies, the identification of the products was possible only on the basis of careful, synchronous analysis of IR and mass spectra.

In the present case, the precise line positions of the measured spectra can be directly compared to literature Ar-matrix IR spectral data, band deconvolution and the analysis of mass spectra are not required for unambiguous identification.

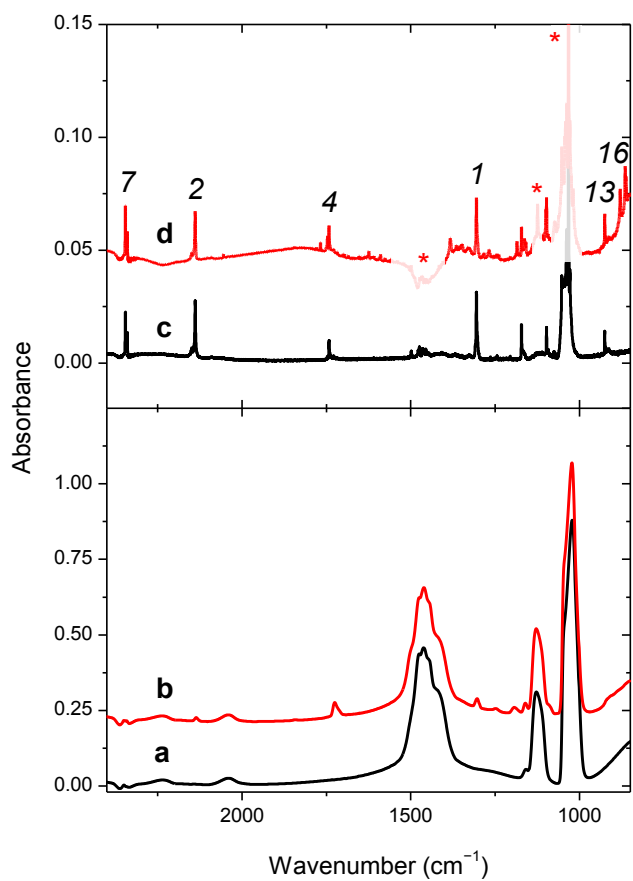

**FIG. 13.** MIR spectrum of the (**a**) neat $CH_3OH$ ice deposited at 12 K on the 4-kelvin substrate; (**b**) the $CH_3OH$ ice after irradiating it with H Lyman-$\alpha$ for 758 min; (**c**) the sample redeposited in an Ar matrix on the substrate of the 10-kelvin cryostat during the TPD process on the 4-kelvin cryostat between 35 and 140 K; (**d**) the difference spectrum obtained by subtracting spectrum recorded at 160 K of the TPD process from the spectrum recorded at the end of the TPD process, when the substrate of the 4-kelvin cryostat was at 300 K. The spectra are shown in the range of 2400–850 $cm^{-1}$. The spectra (**b**) and (**d**) are offset for clarity, the shaded areas marked with an asterisk in (**c**) and (**d**) mask the uncompensated bands of the $CH_3OH$ precursor. See the numerical labelling of the compounds assigned to the most intense bands in Table 1.

For another test case the electron irradiation of the $NH_3 : O_2$ ice, investigated recently by Tsegaw et al., was chosen.[70] In Ref [70], besides nitrogen oxides, hydroxylamine ($NH_2OH$) was identified as the main product. The identification of the products was based on the IR spectrum of the processed ice, and on the PI-ReTOF mass analysis. For ionization 10.49 eV photons were used. Similar studies on the same ice identified basically the same products by IR spectroscopy of the ice and by electron impact mass spectrometry.[71]

**Table 1.** Species detected after the recondensation of the Lyman-α irradiated methanol ice on the upper cryostat in an argon matrix at 11 K. The species listed here were observed by Yocum et al. (2021) and Bennett et al. (2007) after irradiating methanol ices.[67,69]

| Label | Species | Formula [a] | Observed vibrational frequencies (cm$^{-1}$) [b] | Literature value (cm$^{-1}$) | References |
|---|---|---|---|---|---|
| *1* | methane | $CH_4$ | 1305.8 | 1306.0 | [72] |
| *2* | carbon monoxide | CO | 2148.8, 2138.4 | 2148.8, 2138.0 | [73] |
| *3* | formyl radical | HCO | – | 1862.7, 1086.5 | [74] |
| *4* | formaldehyde | $H_2CO$ | 1742.1, 1499.1 | 1742.0, 1498.8 | [75,76] |
| *5* | hydroxymethyl radical | $CH_2OH$ | – | 1183 [c] | [77] |
| *6* | methoxy radical | $CH_3O$ | – | $1044 \pm 2$ [c] | [78] |
| *7* | carbon dioxide | $CO_2$ | 2344.9, 2339.0 | 2345.0, 2339.1 | [79,80] |
| *8* | ketene | $H_2CCO$? | 2142sh? | 2142 [c] | [81] |
| *9* | ethylene oxide | $(CH_2)_2O$? | 879.9*? | 879.0 [c] | [82] |
| *10* | acetaldehyde | $CH_3CHO$ | 1729b, 1348.3? | 1729, 1349 | [83] |
| *11* | vinyl alcohol | $CH_2CHOH$ | – | 818.5 [c] | [82] |
| *12* | ethanol | $CH_3CH_2OH$ | – | 1240.3, 1025.0 | [65] |
| *13* | dimethyl ether | $H_3COCH_3$ | 1455.3?, 1172.4, 1098.2*, 926.0 | 1455.3, 1172.0, 1098.3, 926.1 | [84] |
| *14* | glycolaldehyde | $HOCH_2CHO$ | – | 1746.5, 1110.0 | [85] |
| *15* | methyl formate | $HCOOCH_3$ | 1746.7, 1205.7, 1163.3*, 1159.4* | 1747.1, 1745.2, 1205.6, 1162.5, 1158.1 | [86] |
| *16* | ethylene glycol | $HO(CH_2)_2OH$ | 1163.3*, 1159.4*, 1100.0, 1098.2*, 879.9*, 865.3, 862.9 | 1163.0, 1160.0, 1100.0, 1099.0, 880, 865, 863 | [87] |
| *17* | acetone | $(CH_3)_2CO$ | – | 1721.4, 1361.6, 1216.6 | [63] |

[a] ?: tentative assignment

[b] b:broad; sh: shoulder; ?: uncertain assignment, *: band can be assigned to multiple species due to likely overlapping

[c] only the most intense vibrational bands are listed

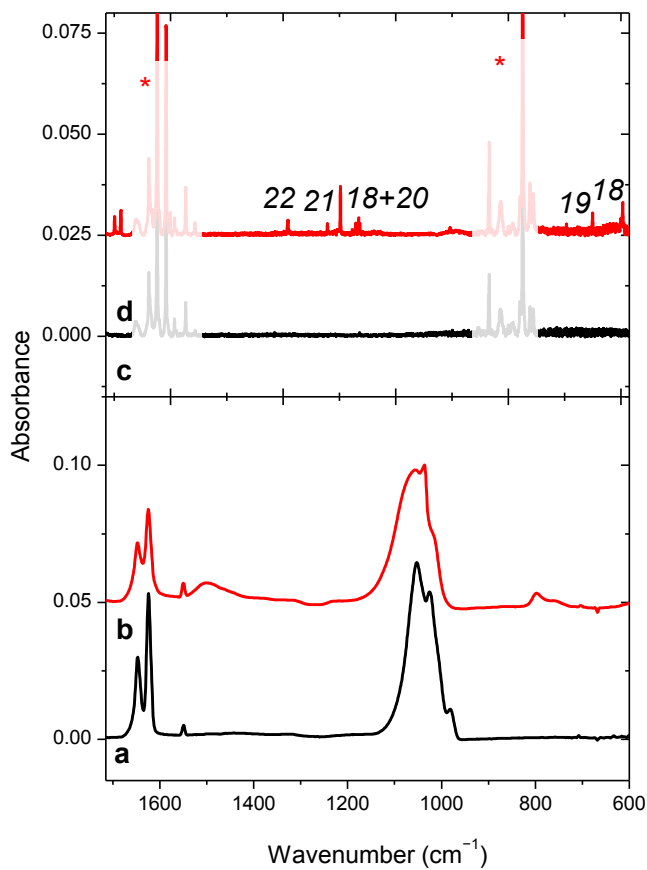

**FIG. 14.** MIR spectrum of the (**a**) $NH_3$ : $O_2$ 2 : 1 ice deposited at 5.5 K on the substrate of the 4-kelvin cryostat; (**b**) the ice mixture after irradiating it with 5 keV electrons for 30 min (emission current was 30 nA); (**c**) the spectrum of the sample redeposited in Ar matrix on the substrate of the 10-kelvin cryostat during the TPD process on the 4-kelvin cryostat between 140 and 190 K; (**d**) and during the TPD process on the 4-kelvin cryostat between 140 and 260 K; the spectral range of 1715–785 cm$^{-1}$. The spectra (**b**) and (**d**) are offset for clarity, the shaded areas marked with an asterisk in (**c**) and (**d**) mask the bands of the $NH_3$ precursor and $H_2O$ irradiation product. See the numerical labelling of the compounds assigned to the most intense bands in the main text.

In our case, to accumulate more products for the TPD – matrix-isolation redeposition FT-IR analysis, 6 deposition (*ca.* 500 nm for the first layer, ~200 nm in the case of the next layers) – electron irradiation cycles (30 min each) were repeated. Here we do not report the full analysis of the redeposited, matrix-isolated products, only the most interesting and exciting results are shown. The lower panel of Fig. 14 depicts the IR spectrum recorded after depositing the first layer of the ice together with the spectrum recorded after electron irradiation. (The same spectral changes were observed for the additional ice deposition – electron irradiation cycles.) The upper panel of Fig. 14 shows the IR spectrum of the sample sublimed between 150 and 240 K and redeposited in an Ar matrix on the substrate of the 10-kelvin cryostat at 11 K. After a short literature search it was easy to identify unambiguously *t*-HONO (*18* in Fig. 14) at 3568.4, 1688.1, 1265.8, 1263.8, 800.6, and 796.6 cm$^{-1}$ (*vs.* literature data at 3568.5, 1688.0, 1265.8, 1263.9, 800.4, and 796.6 cm$^{-1}$),[88] *c*-HONO (*19*) at 1632.8 and 850.2 cm$^{-1}$ (*vs.* literature data at 1632.8 and 850.2 cm$^{-1}$),[88] $H_2O_2$ (*20*) at 1277.5 and 1270.9 cm$^{-1}$ (*vs.* literature data at 1277.0 and 1270.9 cm$^{-1}$),[89] $HNO_3$ (*21*) at 1699.4 and 1321.2 cm$^{-1}$ (*vs.* literature data at 1699.4 and 1321.4 cm$^{-1}$),[90] and *c*-HOONO (*22*) at 3302.4, 1600.8, 1391.5 cm$^{-1}$ (*vs.* literature data at 3303±1, 1600.6±0.6, and 1392±1 cm$^{-1}$).[91] It is worth noting that two bands at 1235 and 796 cm$^{-1}$ can be clearly seen in the spectrum of the processed ice after the evaporation of $O_2$, both in our spectrum and that in Ref. [70]. And although these two bands were assigned to a different species (*i.e.*, to $N_2O_2$) before, based on their positions, they could possibly belong to *t*-HONO. These molecules could not be seen in the PI-ReTOF mass spectra, since they have an ionization energy higher than the applied 10.49 eV photon energy.[70] While upon electron impact ionization they yield the same fragments as nitrogen oxides and hydroxylamine. Therefore, the present example proves that the TPD – matrix-isolation redeposition analysis is a powerful alternative and complementary tool to other analytical methods in the analysis of processed astrophysical analogue ices.

## IV. ASTROCHEMICAL POTENTIAL

Now it is well-accepted by astrochemists that without the consideration and investigation of the low-temperature solid-state chemical processes, taking place on the surface of icy grains, the composition of the gaseous interstellar matter cannot be understood. Although there are many setups constructed for understanding the chemistry induced by different particles and irradiations, it is usually challenging to identify all the molecules produced in the ice and desorbing at a certain temperature to the gaseous phase. This is mostly due to the low concentration of the products and the short time window offered by the TPD process for spectroscopic analysis. A more sensitive frequently used analysis method, the electron impact ionization mass spectrometry has the disadvantage that several molecules can have similar fragmentation patterns or the parent ion can be completely missing from the spectrum due to fragmentation. A solution for this problem was developed by the group of Kaiser,[11] who are using photoionization with very sensitive mass detection. The TPD – matrix-isolation redeposition FT-IR spectral analysis, available by VIZSLA and discussed in this paper as an alternative and, as it is demonstrated, a complementary method to solve this problem and help the analysis of the molecules desorbing from the ice.

VIZSLA is also suitable to study reactions of astrochemical molecules with H atoms and $H^+$ ions in *para*-$H_2$. Although solid *para*-$H_2$ is not a typical astrophysical environment, as it is colder and less polar than a typical astrophysical ice, it provides excellent conditions for very sensitive monitoring of these processes for first glance. VIZSLA has the unique opportunity, that the processes observed and identified in solid *para*-$H_2$, can be studied further with the same setup in astrophysically more relevant ice conditions. The versatile irradiation and particle sources available on VIZSLA, are also important in a view, that the effect of different irradiations can be compared under the same conditions. It offers the unique opportunity to compare the outcome of different type of experiments as well, that otherwise would be difficult to do when the results were obtained in different setups under different conditions.

Besides understanding solid-state astrophysical processes, VIZSLA can provide important spectral and chemical information for flag-ship missions of ESA and NASA. These agencies scheduled the launch of James Webb Space Telescope (JWST) for late 2021.[5] This mission will detect the interstellar matter in the infrared region with unprecedented spectral and special resolution. As it is expected, it will provide spectral information not only on the gaseous interstellar matter, but also on icy dust grains, and maybe on atmospheres of exoplanets. Jupiter Icy Moons Explorer (JUICE), an upcoming mission of ESA will provide chemical information on other, closer icy objects.[92] To understand the data collected by these space missions, extensive laboratory astrochemistry research is required, in which VIZSLA can have a great contribution. Furthermore, the analysis of the molecules desorbing from the ices can provide a valuable list of molecules that can be potentially present in the gaseous phase of ISM and can be searched for example by the Atacama Large Millimeter Array (ALMA) telescope.[93]

It is planned that the different irradiation sources and the sensitivity of VIZSLA will be further calibrated against other ice setups. It is also under consideration that, in part-time, VIZSLA will be offered for collaborations under the EUROPLANET organization.[94]

**DATA AVAILABILITY STATEMENT.**

Data available on request from the authors.

## ACKNOWLEDGEMENTS

The support of the Lendület program of the Hungarian Academy of Sciences is acknowledged. This work was supported by the ELTE Institutional Excellence Program (TKP2020-IKA-05). The authors acknowledge György Oláh and István Gál (Goodwill-Trade Ltd.) for the discussions on the design of the setup and the construction of non-standard vacuum and mechanical parts of the setup. The authors acknowledge Prof. Takamasa Momose (University of British Columbia, Canada), Prof. Yuan-Pern Lee (National Yang Ming Chiao Tung University, Taiwan), Dr. Karolina Haupa (Karlsruhe Institute of Technology, Germany), Prof. Harold Linnartz (Leiden University, The Netherlands), Dr. Jiao He (Max Planck Institute for Astronomy, Germany), and Dr. Marko Förstel (TU Berlin, Germany) for discussions and advices.